\documentclass{ws-procs10x7}

\usepackage{epsfig}

\newcommand{\be}{\begin{eqnarray}}
\newcommand{\ee}{\end{eqnarray}}

\begin{document}

\title{
Implication of the weak phase $\beta$ measured 
in $B \to \rho \gamma$ decay
}

\author{C. S. Kim}

\address{
Department of Physics,
Yonsei University, Seoul 120-749, Korea \\
E-mail: cskim@yonsei.ac.kr }

\author{Yeong  Gyun Kim}

\address{
Department of Physics,
Korea University, Seoul 136-701, Korea\\
E-mail: yg-kim@korea.ac.kr }

\author{Kang Young Lee
\footnote{Conference speaker}
}

\address{
Department of Physics,
KAIST, Daejeon 305-701, Korea\\
E-mail: kylee@muon.kaist.ac.kr }

\twocolumn[\maketitle\abstract{
We explore the exclusive $ B^0 \to \rho^0 \gamma$ decay
to obtain the time-dependent CP asymmetry in $b \to d \gamma$ decay process.
We find that the complex RL and RR mass insertion
to the squark sector in the MSSM
can lead to a large deviation of CP asymmetry from 
that predicted in the Standard Model. 
}]

\section{Introduction}

In the $B$ meson system, 
it is strongly required to find a new observables for
the CP violation in a way independent of the
$B^0$--$\bar{B}^0$ mixing since the observed CP violating asymmetry 
appears only through the mixing so far.
Moreover, we may expect that new physics can influence the $\Delta
B = 1$ penguin decays in a different way from the $\Delta B =2$
mixing, e.g. the controversial deviation of the recent 
measurement of $\sin 2 \beta$ in $B \to \phi K$ decay from
that in $B \to J/\psi K_S$ decays \cite{sin2beta}, 
which implies an evidence of a new physics effect beyond the SM
\cite{DKS}. 

The Cabibbo-suppressed $b \to d \gamma$ decay 
provides us a new chance to study the CP violation 
in a way independent of the mixing. 
In the present work, we consider the time-dependent CP asymmetry
in the neutral $B^0 \to \rho^0 \gamma$ decay.
Although we will be able to determine $V_{td}$ 
from the inclusive $B \to X_d \gamma$ decay
in a theoretically clean way \cite{KSM}, 
it suffers from large $B \to X_s \gamma$ background in the experiment. 
The charged $B^\pm \to \rho^\pm
\gamma$ decay mode provides clean signal and has a branching ratio
twice larger than that of the neutral mode, by the isospin
symmetry. However, the long-distance (LD) effect on the charged
mode due to dominantly $W^\pm$-annihilation is very large
($\sim$30 \%), which contaminates the CP violating effect
\cite{alibraun,wyler}.
The exclusive $B \to \rho \gamma$ decays in the SM
and the MSSM have been studied in the literature
\cite{alihandoko}. 

The photon has two helicity states $\gamma_L$ and $\gamma_R$
although we cannot discriminate them in the experiment.
Since the time-dependent CP violating asymmetry is defined 
when both $B$ and $\bar B$ mesons decay into a same state, 
there is no interference between final states
with the definite helicity.
In the SM, the operator which governs $b \to d \gamma$ decay is chiral
and the conjugate operator is suppressed by $m_d/m_b$
and the CP asymmetry also suppressed accordingly.
Therefore the new physics beyond the SM is required for a large
time-dependent CP asymmetry enough being observed
in the experiment \cite{soni}.

In this work, we consider the supersymmetric models which have
non-diagonal elements of the squark mass matrices,
parameterized by the mass insertions
$(\delta_{ij})_{MN} \equiv (\tilde{m}_{ij}^2)_{MN} /{\tilde{m}}^2 $,
where $\tilde{m}$ is the averaged squark mass,
$i$ and $j$ are flavor indices and $M$ and $N$ denote
chiralities.
The $\delta$'s are complex in general and provide new CP phases.
To simplify our discussion,
we consider only $(\delta_{13})_{RL}$ and $(\delta_{13})_{RR}$
dominating cases.
In section 2, we describe the $B^0 \to \rho^0 \gamma$ decay
and the time-dependent CP asymmetry.
The supersymmetric contributions are given in section 3
and the numerical results given in the section 4.
We conclude in section 5.

\section{CP asymmetry in $B^0 \to \rho^0 \gamma$ decay}

The relevant terms of the effective Hamiltonian
for the $b \to d \gamma$ decay is written as
\be
{\cal H}_{\rm eff} &=& \frac{4 G_F}{\sqrt{2}}
            \sum_{q=u,c} \left[ \lambda_q
            \sum_{i=1,2} 
                 \left( C_i O_i^q + C'_i {O'_i}^q \right) \right.,
\nonumber \\
&&
\left.
   - \lambda_t \left( C^{\rm eff}_7 O_7
                 + C^{\prime \rm eff}_7 O'_7 \right)
            + \cdot \cdot \cdot \right],
\ee
where $\lambda_q = V_{qb} V_{qd}^*$,
$O_1^q = ({\bar{d}_L}^\alpha \gamma_\mu {q_L}^\beta)
        ({\bar{q}_L}^\beta \gamma^\mu {b_L}^\alpha)$,
$O_2^q = (\bar{d}_L \gamma_\mu q_L) 
             (\bar{q}_L \gamma^\mu  b_L)$,
and
$O_7 = (e m_b/16 \pi^2) \bar{d}_L \sigma_{\mu \nu}
                                  F^{\mu \nu}  b_R$.
The primed $O'_i$ are their chiral conjugate operators.
The effective Wilson coefficient $ C^{(\prime) \rm eff}_7$ includes
the effects of operator mixing.

We write the amplitudes for the final states of polarized photon as
\be
A_L &\equiv& \langle \rho \gamma_L | H_{\rm eff} | B^0 \rangle
\sim {C^{\prime \rm eff}_7}^* \lambda_t^*
 \langle \rho \gamma_L | {O'_7}^\dagger | B^0 \rangle,
\nonumber \\
A_R &\equiv& \langle \rho \gamma_R | H_{\rm eff} | B^0 \rangle
\sim {C^{\rm eff}_7}^* \lambda_t^*
 \langle \rho \gamma_R | {O_7}^\dagger | B^0 \rangle,
\nonumber \\
\bar{A}_L &\equiv& \langle \rho \gamma_L | H_{\rm eff} | \bar{B}^0 \rangle
\sim C^{\rm eff}_7 \lambda_t
 \langle \rho \gamma_L | O_7 | \bar{B}^0 \rangle,
\nonumber \\
\bar{A}_R &\equiv& \langle \rho \gamma_R | H_{\rm eff} | \bar{B}^0 \rangle
\sim C^{\prime \rm eff}_7 \lambda_t
 \langle \rho \gamma_R | O'_7 | \bar{B}^0 \rangle,
\ee
up to the factor of $4 G_F/\sqrt{2}$.
We note that
$\langle \rho \gamma_L | O_7 | \bar{B}^0 \rangle 
= \langle \rho \gamma_L | {O'_7}^\dagger | B^0 \rangle$,
and $ \langle \rho \gamma_R | O'_7 | \bar{B}^0 \rangle
= \langle \rho \gamma_R | {O_7}^\dagger | B^0 \rangle$.
In the SM, $ C^{\prime \rm eff}_7$ is suppressed by the mass ratio
$m_d/m_b$ and so is the right polarized photon emission
$b_L \to q_R \gamma_R$.
For the neutral $B$ meson decay, the LD contribution
due to $W$-exchange is merely a few \% from the QCD sum rule
calculation \cite{alibraun,wyler}, so it will be ignored in our analysis.
We investigate the time-dependent CP asymmetry given by
\be
A_{\rm CP}(t) &=&
\frac{\bar{\Gamma} - \Gamma}
     {\bar{\Gamma} + \Gamma}
\\
&\equiv& - {\cal C} \cos (\Delta m_B t) + {\cal S} \sin (\Delta m_B t),
\nonumber
\ee
where
$\bar{\Gamma} = \Gamma(\bar{B}^0(t) \to \rho^0 \gamma_L)
      + \Gamma(\bar{B}^0(t) \to \rho^0 \gamma_R)$,
$\Gamma = \Gamma(B^0(t) \to \rho^0 \gamma_L)
      + \Gamma(B^0(t) \to \rho^0 \gamma_R)$,
since we cannot distinguish $\gamma_L$ and $\gamma_R$ in practice.
The coefficients 
${\cal C} = 0$ and
\be
{\cal S} = \frac{ |A_L|^2 {\rm Im} \lambda_L + |A_R|^2 {\rm Im} \lambda_R}
           {|A_L|^2 + |A_R|^2},
\ee
with the parameter $\lambda_{L(R)}$ defined by
\be
\lambda_{L(R)} \equiv \sqrt{\frac{M_{12}^*}{M_{12}}}
                      \frac{\bar{A}_{L(R)}}{A_{L(R)}}.
\ee
The off-diagonal element $M_{12}$ describes the $B^0$--$\bar{B}^0$
mixing and $A_{L(R)}$ does the $b \to d \gamma$ decays.
We define $ 2 \beta_{\rm mix} = {\rm Arg} \left( M_{12} \right)$
and $ 2 \beta_{\rm decay} =  {\rm Arg} \left( {\bar A}_R/A_R \right)
 = {\rm Arg} \left( {\bar A}_L/A_L \right)$.
Then the coefficient ${\cal S}$ is expressed by
\be
{\cal S} = - \frac{2~|C_7||C'_7|}{|C_7|^2+|C'_7|^2}
          \sin (2 \beta_{\rm mix} - 2 \beta_{\rm decay}),
\ee
where we rewrite
\be
2 \beta_{\rm decay} = 2 \beta_{\rm SM} + {\rm Arg}(C'_7) -  {\rm Arg}(C_7^*).
\ee
Note that we have an additional factor
$2~|C_7||C'_7|/(|C_7|^2+|C'_7|^2)$,
which can enhance or suppress ${\cal S}$
by the new physics effect $|C'_7|$.

\section{SUSY contributions}

By penguin diagrams with gluino-squark loop,
the Wilson coefficients $C'_i$ 
get contribution to produce $\gamma_R$ at the matching scale $\mu = m_W$.
After the RG evolution, we have $ C^{\rm eff}_7(m_b) = C^{\rm SM}_7(m_b)
= -0.31$ and
\be
C^{\prime \rm eff}_7(m_b) &=&  \frac{\sqrt{2}}{G_F V_{tb} V_{td}^*}
       \left( 0.67~C^{\rm SUSY}_7(m_W) \right.
\nonumber \\
 && ~~ \left.+ 0.09~C^{\rm SUSY}_8(m_W) \right),
\ee
where the SUSY contributions at $\mu = m_W$ are
\be
&& C^{\rm SUSY}_7 = \frac{4 \alpha_s \pi Q_b}{3{\tilde m}^2}
         \left[ (\delta_{13})_{RR} M_4 (x) \right.
\nonumber \\
&&~~~~~~~~~~~ \left. - (\delta_{13})_{RL} 4 B_1(x) 
                       \frac{m_{\tilde g}}{m_b} \right],
\\
&& C^{\rm SUSY}_8 = \frac{\alpha_s \pi}{6 {\tilde m}^2}
         \left[ (\delta_{13})_{RR} (9 M_3(x) - M_4(x) ) \right.
\nonumber \\
&&~~~~ \left. + (\delta_{13})_{RL} \left(4 B_1(x)-9 \frac{B_2(x)}{x} \right)
                \frac{m_{\tilde g}}{m_b} \right],
\nonumber
\ee
with 
$x=(m_{\tilde{g}}/\tilde{m})^2$
\cite{everett}.
Note that the SUSY contribution is more sensitive to $(\delta_{13})_{RL}$
than $(\delta_{13})_{RR}$ due to the enhancement factor $m_{\tilde g}/m_b$.
The loop functions $B_i(x)$ are found in the literature
\cite{everett}.
Since $\delta_{RL,RR}$ are complex in general,
the Wilson coefficients ${C'}_7^{\rm eff}(m_b)$ has nontrivial phase
which affects the phase of $\bar{A}/A$.

On the other hand, the $B$--$\bar{B}$ mixing is  affected
by the gluino-squark box diagrams in the MSSM.
The relevant  $\Delta B =2$ effective Hamiltonian
with the supersymmetric contribution contains new 
scalar-scalar interaction operators
$O'_{S2} = (\bar{d}_\alpha (1+\gamma_5) b_\alpha)
      (\bar{d}_\beta  (1+\gamma_5) b_\beta)$,
$O'_{S3} = (\bar{d}_\alpha (1+\gamma_5) b_\beta)
      (\bar{d}_\alpha  (1+\gamma_5) b_\beta)$,
when we introduce only the RL and RR mass insertions.
The Wilson coefficient $C_1$
corresponding to the SM operator $O_1 = (\bar{d} \gamma_\mu (1-\gamma_5) b)
(\bar{d} \gamma_\mu (1-\gamma_5) b)$ consists of
the SM part and the supersymmetric contributions,
while $C'_{S2}$ and $C'_{S3}$
corresponding to the above operators are entirely supersymmetric.
Their explicit expression at the scale $\mu = M_{\rm SUSY}$
can be found in Refs. \cite{gabrielli,ko}.
The RG evolved Wilson coefficients from  $m_W$ to $m_b$ scale
ignoring the RG running effects between $M_{\rm SUSY}$ and $m_W$,
are given 
in  Ref. \cite{BB}.

\section{Numerical results}

Figure 1 shows the quantity ${\cal S}$
as a function of the phase of $(\delta_{13})_{RL}$, $\varphi$,
assuming $|(\delta_{13})_{RL}|=0.001$.
We vary the weak phase $\gamma$ from 0 to $ 2\pi$.
Hereafter we use the input parameters as follows:
$ m_B= 5.3~{\rm GeV}$, $ m_t= 174.3~{\rm GeV}$, $ m_b= 4.6~{\rm GeV}$,
and $ \alpha_s(m_Z) = 0.118$.
The decay constant $f_{B_d} = 200 \pm 30$ MeV is
the main source of the theoretical uncertainty
and the bag parameters are those of  Ref. \cite{bag};
$B_1 = 0.87,~~ B_2=0.82,~~ B_3=1.02$.
The supersymmetric scale is taken to be
$m_{\tilde{g}} \approx \tilde{m} \approx M_{\rm SUSY} \approx 500$ GeV.
We require that the mass difference $\Delta m_B$
and $\beta_{\rm mix}$ in $B \to J/\psi K$ decay
should be within the experimental limit:
$\Delta m_B = 0.489 \pm 0.008$ ps$^{-1}$ \cite{pdb}
and $\sin 2 \beta_{\rm mix} = 0.734 \pm 0.055$ \cite{sin2beta}.
We do not use  ${\rm Br}(B \to \rho / \omega \gamma)$
as a constraint since it involves a large theoretical uncertainty
in the form factor.
Instead, we assume a moderate upper bound on
the branching ratio of the inclusive $B \to X_d \gamma$ decay
${\rm Br}(B \to X_d \gamma) < 1.0 \times 10^{-5}$,
following Ref. \cite{ko}.
although the inclusive decay is not observed yet.
The black region corresponds to the allowed values for
the phase of $(\delta_{13})_{RL}$,
while the grey (green) region denotes the parameter set which satisfies
the $\Delta m_B$ and $\sin 2 \beta_{\rm mix}$ constraints
but exceeds the bound on ${\rm Br}(B \to X_d \gamma)$.
We find that large CP violating asymmetry is possible.

The plot of ${\cal S} $ with respect to $|(\delta_{13})_{RL}|$
is depicted in  Fig. 2
when the phase $\varphi$ is fixed to be zero.
The black region and the grey (green) region
are defined as  in Fig. 1.
We see that $|(\delta_{13})_{RL}|$ is strongly constrained by the
inclusive branching ratio and a large CP violation is still possible
even when ${C'}_7^{\rm eff}(m_b)$ is real.
The branching ratio ${\rm Br}(B \to X_d \gamma)$
and CP asymmetry ${\cal S}$ provide the complimentary information
on  $(\delta_{13})_{RL}$.

\begin{figure}
\epsfxsize160pt
\figurebox{160pt}{160pt}{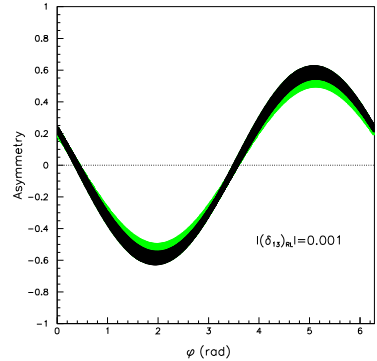}
\caption{
The time-dependent CP asymmetry ${\cal S}$
as a function of the phase of $(\delta_{13})_{RL}$.
$|(\delta_{13})_{RL}| =0.001$ is assumed.
The black region denotes allowed points while grey (green) region
excluded points by the inclusive $b \to d \gamma$
branching ratio bound.
}
\label{fig:f1}
\end{figure}

\begin{figure}
\epsfxsize160pt
\figurebox{160pt}{160pt}{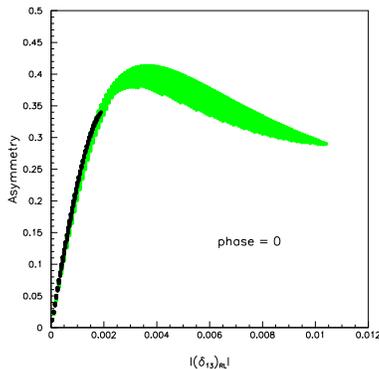}
\caption{
The time-dependent CP asymmetry ${\cal S}$
as a function of $|(\delta_{13})_{RL}|$.
The phase of $(\delta_{13})_{RL}$ is assumed to be 0.
The black region and the grey (green) region
are defined in Fig. 1.
}
\label{fig:f2}
\end{figure}

\section{Concluding remarks}

If we observe a sizable CP asymmetry in $B^0 \to \rho^0 \gamma$ decay,
it will be a clear evidence of the new physics
beyond the SM.
Although it is hardly expected that the time dependent CP asymmetry
of $B^0 \to \rho^0 \gamma$ will be measured in the present $B$-factory,
it will be achieved in the next generation of $B$-factory
with about 100 times more $B$ mesons produced.
Due to the agreement of the SM prediction with
the present $\Delta m_B$ data and the CP asymmetry
in $B \to J/\psi K$ decay,
we favor the new physics which contributes less
to the $B$--$\bar{B}$ mixing
but has a strong  effect on the $b \to d \gamma$
penguin diagram.
In this work, we showed that the RL mass insertion of squark mixing
of the MSSM can produce a large CP asymmetry
of $B^0 \to \rho^0 \gamma$ decay process.

\def\PRD #1 #2 #3 {Phys. Rev. D {\bf#1},\ #2 (#3)}
\def\PRL #1 #2 #3 {Phys. Rev. Lett. {\bf#1},\ #2 (#3)}
\def\PLB #1 #2 #3 {Phys. Lett. B {\bf#1},\ #2 (#3)}
\def\NPB #1 #2 #3 {Nucl. Phys. {\bf B#1},\ #2 (#3)}
\def\ZPC #1 #2 #3 {Z. Phys. C {\bf#1},\ #2 (#3)}
\def\EPJ #1 #2 #3 {Euro. Phys. J. C {\bf#1},\ #2 (#3)}
\def\JHEP #1 #2 #3 {JHEP {\bf#1},\ #2 (#3)}
\def\NC #1 #2 #3 {Nuovo Cimento {\bf#1A},\ #2 (#3)}
\def\IJMP #1 #2 #3 {Int. J. Mod. Phys. A {\bf#1},\ #2 (#3)}
\def\MPL #1 #2 #3 {Mod. Phys. Lett. A {\bf#1},\ #2 (#3)}
\def\PTP #1 #2 #3 {Prog. Theor. Phys. {\bf#1},\ #2 (#3)}
\def\PR #1 #2 #3 {Phys. Rep. {\bf#1},\ #2 (#3)}
\def\RMP #1 #2 #3 {Rev. Mod. Phys. {\bf#1},\ #2 (#3)}
\def\PRold #1 #2 #3 {Phys. Rev. {\bf#1},\ #2 (#3)}
\def\IBID #1 #2 #3 {{\it ibid.} {\bf#1},\ #2 (#3)}

\end{document}